\documentclass[%
 aip,
 jmp,%
 amsmath,amssymb,
 preprint,%
]{revtex4-1}

\usepackage{dcolumn}
\usepackage{graphicx,amssymb,color}
\usepackage{gensymb}
\usepackage{bm}

\begin{document}

\begin{center}
\noindent\LARGE{\textbf{One-dimensional Silicon and Germanium Nanostructures With No Carbon Analogues}}
\vspace{0.6cm}

\noindent\large{\textbf{E. Perim $^{\ast}$ $^a$, R. Paupitz \textit{$^{b}$}, T. Botari \textit{$^{a}$} and
D. S. Galvao\textit{$^{a}$}}}\vspace{0.5cm}

\textit{$^{a}$ Instituto de F\'isica ‘Gleb Wataghin’, Universidade Estadual de Campinas, 13083-970, Campinas, SP, Brazil.}

\textit{$^{b}$ Departamento de F\'isica - Universidade Estadual
Paulista Av.24A 1515 - 13506-900 - Rio Claro -
SP - Brazil.}

$^{\ast}$ Corresponding author: eric.perim@gmail.com
\end{center}

%Please note that \ast indicates the corresponding author(s) but no footnote text is required. 

%\noindent\textit{\small{\textbf{Received Xth XXXXXXXXXX 20XX, Accepted Xth XXXXXXXXX 20XX\newline
%First published on the web Xth XXXXXXXXXX 200X}}}

%\noindent \textbf{\small{DOI: 10.1039/b000000x}}
%\vspace{0.6cm}
%Please do not change this text.

\noindent \normalsize{In this work we report new silicon and germanium tubular nanostructures with no corresponding stable carbon analogues. The electronic and mechanical properties of these new tubes were investigated through \emph{ab initio} methods. Our results show that the structures have lower energy than their corresponding nanoribbon structures and are stable up to high temperatures (500 and 1000 K, for silicon and germanium tubes, respectively). Both tubes are semiconducting with small indirect band gaps, which can be significantly altered by both compressive and tensile strains. Large bandgap variations of almost $50\%$ were observed for strain rates as small as $3\%$, suggesting possible applications in sensor devices. They also present high Young's modulus values (0.25 and 0.15 TPa, respectively). TEM images were simulated to help the identification of these new structures.}
\vspace{0.5cm}

\section{Introduction}

Carbon nanostructures present very interesting electronic and mechanical properties. The discovery of fullerenes \cite{kroto1985c}, carbon nanotubes \cite{iijima1991helical}, and more recently graphene \cite{novoselov2004electric}, has created a new era in materials science. In particular, the discovery of new and very unusual graphene properties has led to a renewed interest in the search for other similar structures.

A natural question is whether other atoms in the same column of the periodic table as carbon (such as, silicon and germanium) could produce similar structures. This has motivated many studies \cite{rothlisberger1994structure, takeda1994theoretical, zunger1996theory, singh2004metal}, which produced important results. It has been demonstrated that silicon and germanium are able to produce many analogues of the carbon nanostructures, such as closed cage structures \cite{rothlisberger1994structure}, nanotubes \cite{singh2004metal, fagan2000ab, guo2012ab, bai2004metallic, sha2002silicon, zhang2002silicon, durgun2005silicon, pradhan2005hybrid}, and even two-dimensional honeycomb graphene-like sheets \cite{Ref8, cahangirov2009two, topsakalprb, PLA2012, C4CP02902J}, the so-called silicene and germanene,  which have been already experimentally realized \cite{ni2011tunable, o2012stable, vogt2012silicene, davila2014germanene}. More recently, multilayer silicene has been synthesized  and shown to be promising for electronic applications \cite{vogt2014synthesis} as well as stable under ambient conditions \cite{2053-1583-1-2-021003}.

\begin{figure}[h!]
\centerline{\includegraphics[width=8cm]{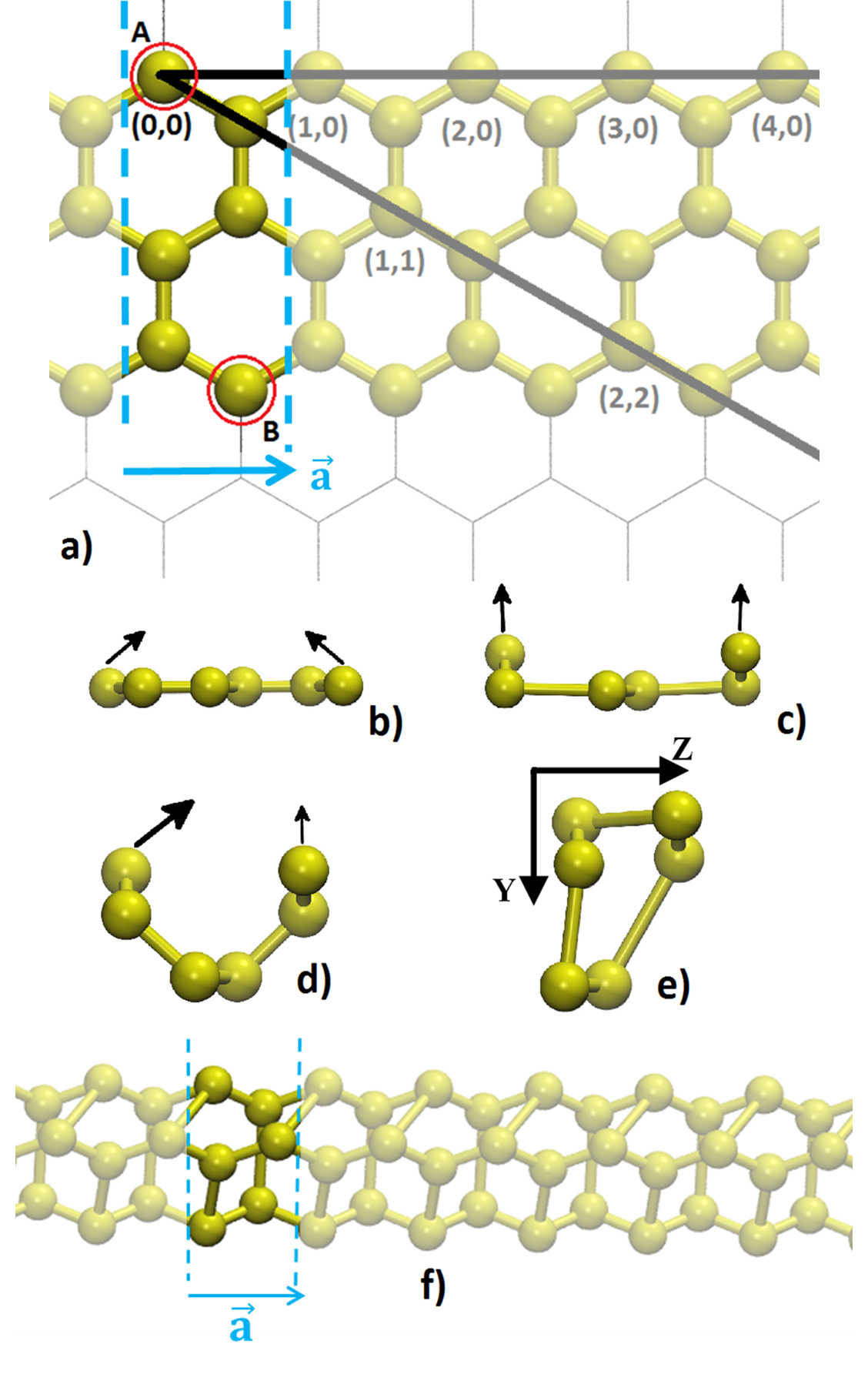}}
\caption{(color online) (a) Top view of the parent nanoribbon. The circled atoms A and B will be connected to form the tube. Atoms belonging to the tube unit cell are highlighted. The vector $\vec{a}$ indicates the tube axis and the lattice parameter. (b), (c), (d) and (e) Side view of intermediate steps of the conversion from the ribbon to the tube. (f) Resulting tube with highlighted unit cell.}
\label{figure1}
\end{figure}

However, despite all these similarities, there are important and significant differences among carbon and silicon and/or germanium structures. Due to the pseudo Jahn-Teller effect (PJTE) \cite{hobey1965vibronic, jose2012understanding, bersuker2013pseudo}, silicon and germanium nanostructures tend to form buckled geometries, as a consequence of a stronger $sp^3$ character, in comparison to carbon ones. Structural buckling of both silicene and germanene structures has been already demonstrated \cite{cahangirov2009two}. In principle, due to these differences, it is possible that unique silicon and germanium structures can exist, with no corresponding carbon counterpart. In this work we report new one-dimensional silicon and germanium nanostructures, with no corresponding stable carbon analogues. Not only these structures break the usual analogy, they are also mechanically robust and present promising electronic properties for technological applications.

\section{Methodology} 
\label{sec2}

The structural, electronic and mechanical properties of these new structures were investigated through \emph{ab initio} density functional theory (DFT) methods using the DMol$ˆ3$ \cite{delley2000dmol3} package, as implemented in the Accelrys Materials Studio Suite. The DFT calculations were carried out in the generalized gradient approximation (GGA), with the Perdew-Burke-Ernzerhof (PBE) \cite{perdew1996generalized} functional for exchange-correlation terms. 
All calculations were carried with double-numeric quality basis set with polarization function and all-electron core treatment. Geometric optimizations were carried out with a tolerance of $10^{-5}$Ha in energy, 0.002 Ha/\AA\ in force and 0.005\AA\ in spatial displacement with full cell parameter optimizations. A k-point grid of 5x1x1 was used, yielding a separation of $0.05$\AA$^{-1}$ between k-points in the reciprocal space. The molecular dynamics (MD) simulations were performed under the Born-Oppenheimer aproximation in a NVT ensemble, with time steps of 1.0fs and massive generalized gaussian moments (GGM) thermostat with Nos\'e chain length of 2 and a Yoshida parameter \cite{delley2000dmol3} of 3 for a simulation time of 5ps.

\section{Results and Discussion}
\label{sec3}
The new proposed silicon and germanium nanotubes are presented in Figure~\ref{figure1}. These are the smallest possible nanotubes which can be formed from a two-ring wide zigzag nanoribbon structure. One unique aspect of these tubes is that they present no helicity despite being neither armchair or zigzag (see Figure~\ref{figure1} (a)). These structures can be generated moving the three atomic chains that compose the nanoribbon into a tubular configuration (see Figure~\ref{figure1}) with six atoms in the unit cell. The transverse dimensions along the directions y and z specified in Figure \ref{figure1} are, respectively, 6.36\AA\ and 4.69\AA\ for the Si tube and 6.98\AA\ and 5.15\AA\ for the Ge tube.

\begin{figure}
\centerline{\includegraphics[width=8cm]{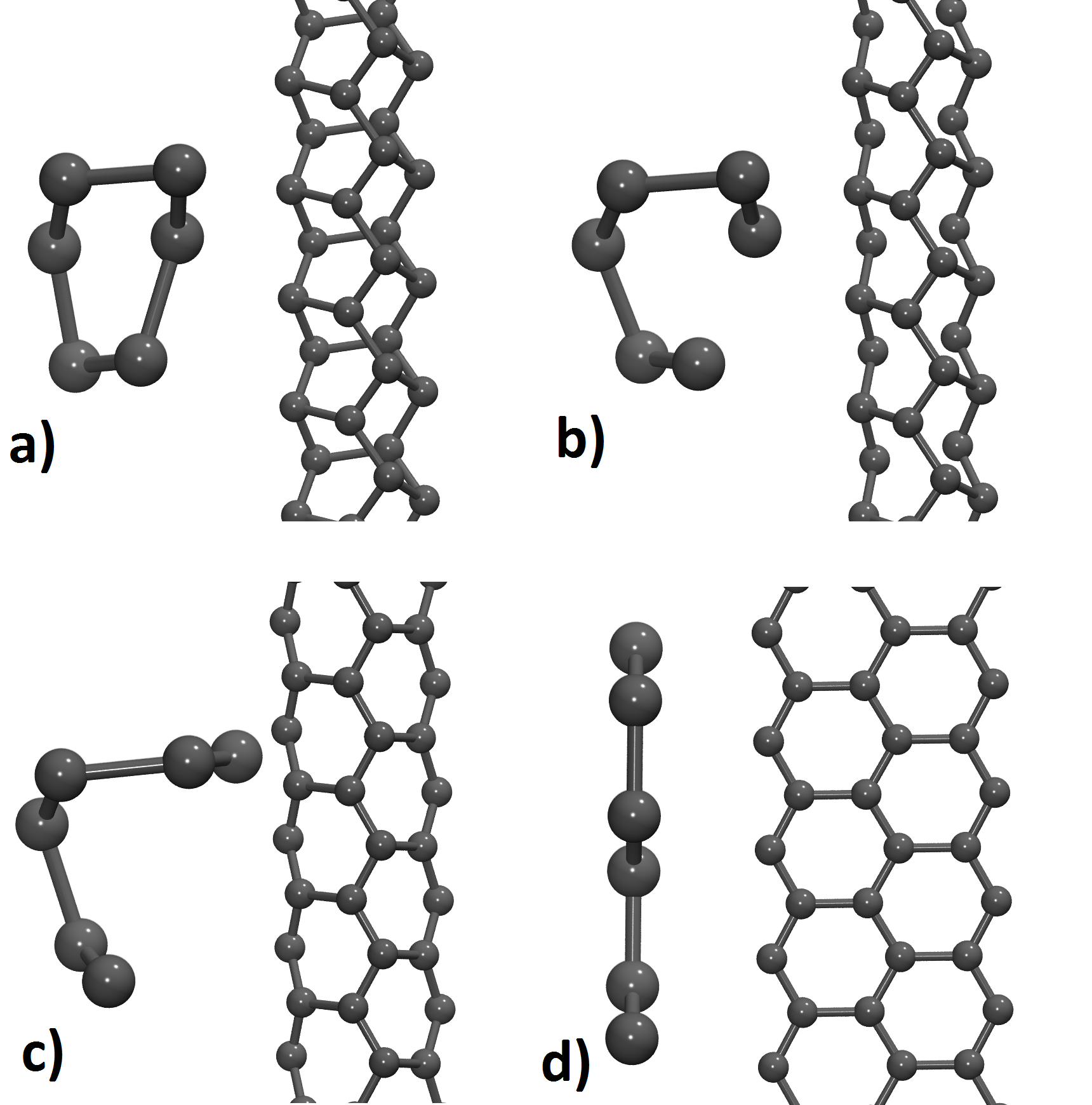}}
\caption{Representative snapshots showing cross-section and front views of the carbon nanotube optimization process. From (a) to (d) we can follow the process in which the structure collapses from an initial tubular (and partially $sp^3$ hybridized) morphology into a planar (and purely $sp^2$ hybridized) nanoribbon.}
\label{COpt}
\end{figure}

No stable carbon structure can be formed with this morphology, the simulations showed it collapses into its planar nanoribbon conformation. Successive snapshots of the carbon nanotube optimization process are shown in Figure \ref{COpt}, depicting the transition from the tubular structure, which has a partial $sp^3$ character, to the planar and purely $sp^2$ nanoribbon. To understand why these silicon and germanium structures are stable while the carbon one is not, we must consider the pseudo Jahn-Teller Effect (PJTE) \cite{hobey1965vibronic, jose2012understanding, bersuker2013pseudo}. Since silicon and germanium $p$ orbitals are much closer in energy than the corresponding carbon ones \cite{jose2012understanding}, the PJTE is much stronger in silicon and germanium structures than on their carbon counterparts. This is reflected into a stronger $sp^3$ hybridization character, which favors and stabilizes the $sp^3$-like tubular structures. In the carbon case, since the $sp^2$ hybridization is favored (and more stable), the small-diameter tubular structures become unstable. The manifestation of the PJTE in these aspects is very interesting, as it breaks the usual structural analogy between carbon and silicon or germanium, leading to the existence of stable nanostructures with no carbon analogues.
The silicon and germanium tubes are thermally stable up to temperatures of 500 and 1000 K, respectively. See Supplemental Material for a video showing successive geometry optimization cycles of the C tube, depicting the transition from the tubular morphology to the nanoribbon. 

\begin{figure}
\centerline{\includegraphics[width=8cm]{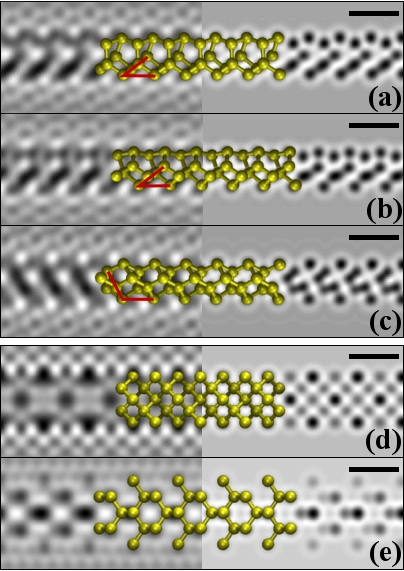}}
\caption{ (color online) Simulated TEM images. (a), (b) and (c) show three different angles of a silicon nanotube. Measured values for the depicted angles are (a) 37\degree; (b) 38\degree; (c) 119\degree; (d) and (e) show a silicon nanowire grown along the (1 0 0) direction and imaged from the (d) (0 1 0); (e) (1 1 0) directions. The left side of the figures show images simulated with spherical aberration of 0.5mm, while the right side images were simulated with spherical aberration of 0.005mm. Scale bars correspond to 6\AA\ .}
\label{TEM}
\end{figure}

In order to facilitate their possible identification, we simulated transmission electron microscopy (TEM) images of the structures and compared them with an image of a very thin silicon nanowire grown along the (1 0 0) direction. This comparison is shown in Figure \ref{TEM}. The TEM images were generated using the QSTEM software \cite{koch2010qstem} with a voltage of 200kV, Scherzer defocus and two different values for the spherical aberration, 0.5mm and 0.005mm, trying to emulate a more conventional microscope and a higher end one, respectively. We speculate that such structures could be observed as a result of experiments similar to those performed by Takayanagi \emph{et al.} \cite{ohnishi1998quantized}, Lou \emph{et al.} \cite{lu2010cold}, Yacaman \emph{et al.} \cite{troiani2003direct} or Ajayan and IIjima \cite{ajayan1992electron}, where strain is used to induce the formation of one-dimensional structures. Considering the recent efforts on the production of silicene and germanene \cite{ni2011tunable, o2012stable, vogt2012silicene}, the successful production of free-standing silicene and germanene layers will greatly increase the chances of the observation of such structures.

\begin{figure}[h!]
\centerline{\includegraphics[width=7cm]{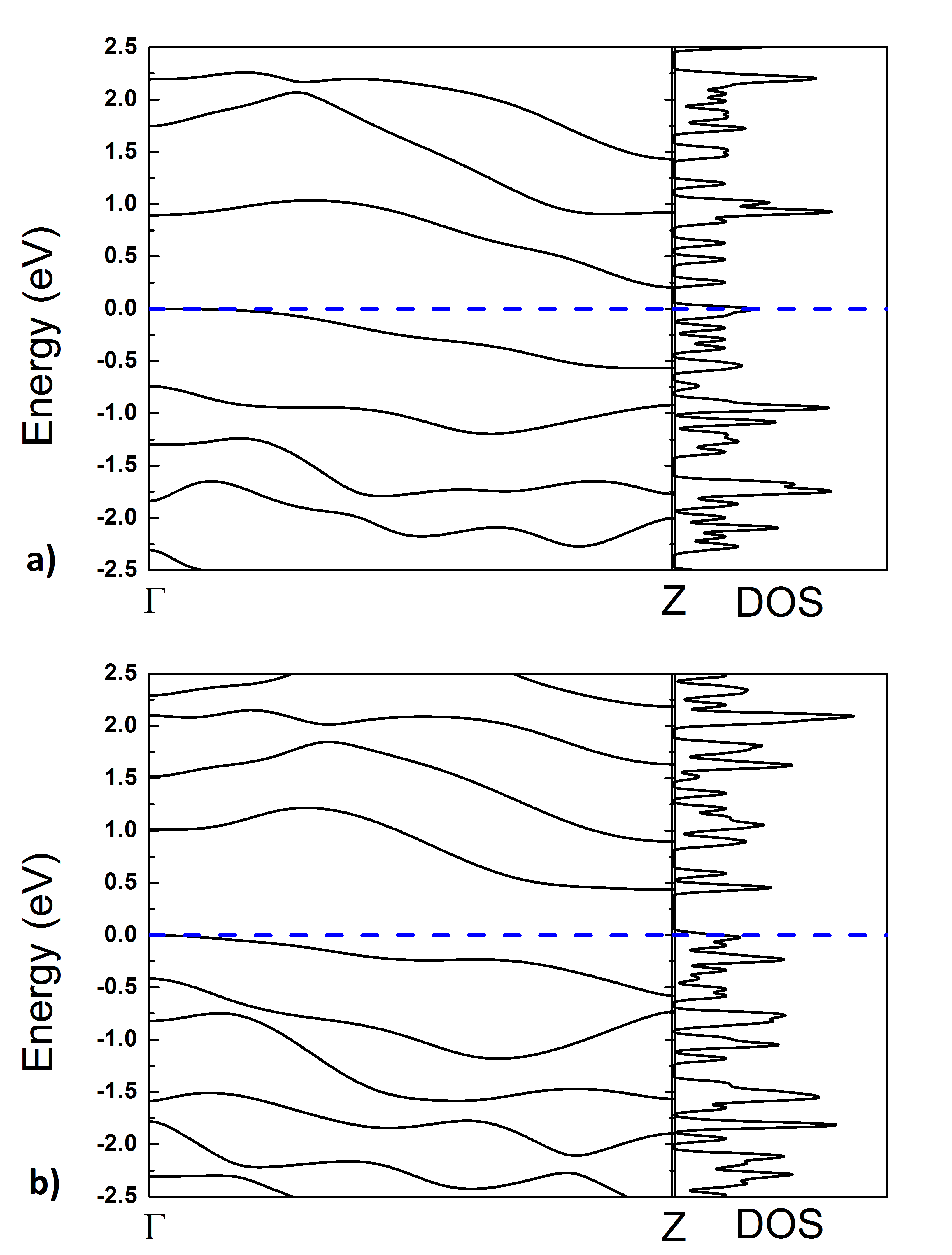}}
\caption{Band structures for (a) silicon and (b) germanium nanotubes. The Fermi level value is indicated by the dashed line.}
\label{bandstructure}
\end{figure}

In Figure ~\ref{bandstructure} we present the band structures and the density of states (DOS) for both silicon and germanium tubes. All structures present bands with significant dispersion and small indirect band gaps. The bandgap values are of 0.44 eV and 0.22 eV for  silicon and germanium structures, respectively. Even considering the fact that DFT calculations usually underestimate bandgap values \cite{perdew1983physical}, these values are reasonably small.

\begin{table*}%[h]
\begin{tabular}{p{0.15\linewidth} p{0.15\linewidth} p{0.10\linewidth} p{0.20\linewidth} p{0.14\linewidth} p{0.10\linewidth}}%{lccccc}
              & Bandgap (eV)             & $a$ (\AA)     & Young Modulus (TPa)       & Poisson Ratio        & $E_{f}$ (eV)    \\ \hline
Si Tube   & 0.22              & 3.777             & 0.25             & 0.12 - 0.12          & -4.02             \\
Si Bulk   & -                    & -                    & 0.08              & -                    & -4.93              \\
SiNR       & -                    & -                    & -                    & -                    & -3.84             \\ \hline
Ge Tube & 0.44              & 4.026            & 0.15              & 0.07 - 0.06          & -3.52              \\
Ge Bulk & -                    & -                    & 0.06              & -                    & -4.14              \\
GeNR      & -                    & -                    & -                    & -                    & -3.29             \\ \hline
               & \multicolumn{1}{l}{} & \multicolumn{1}{l}{} & \multicolumn{1}{l}{} & \multicolumn{1}{l}{} & \multicolumn{1}{l}{}
\end{tabular}
    \caption{Silicon and germanium nanostructures properties. SiNR and GeNR stand for a two-ring wide zigzag silicene and germanium nanoribbons, respectively. $a$ is the lattice parameter and $E_{f}$ is the formation energy per atom. Poisson's ratios are calculated along the directions y/z, as specified in Fig. ~\ref{figure1}. Young's modulus for bulk crystals were calculated along the (1 0 0) direction.}
    \label{table1}
\end{table*}

\begin{figure*}
\centerline{\includegraphics[width=14cm]{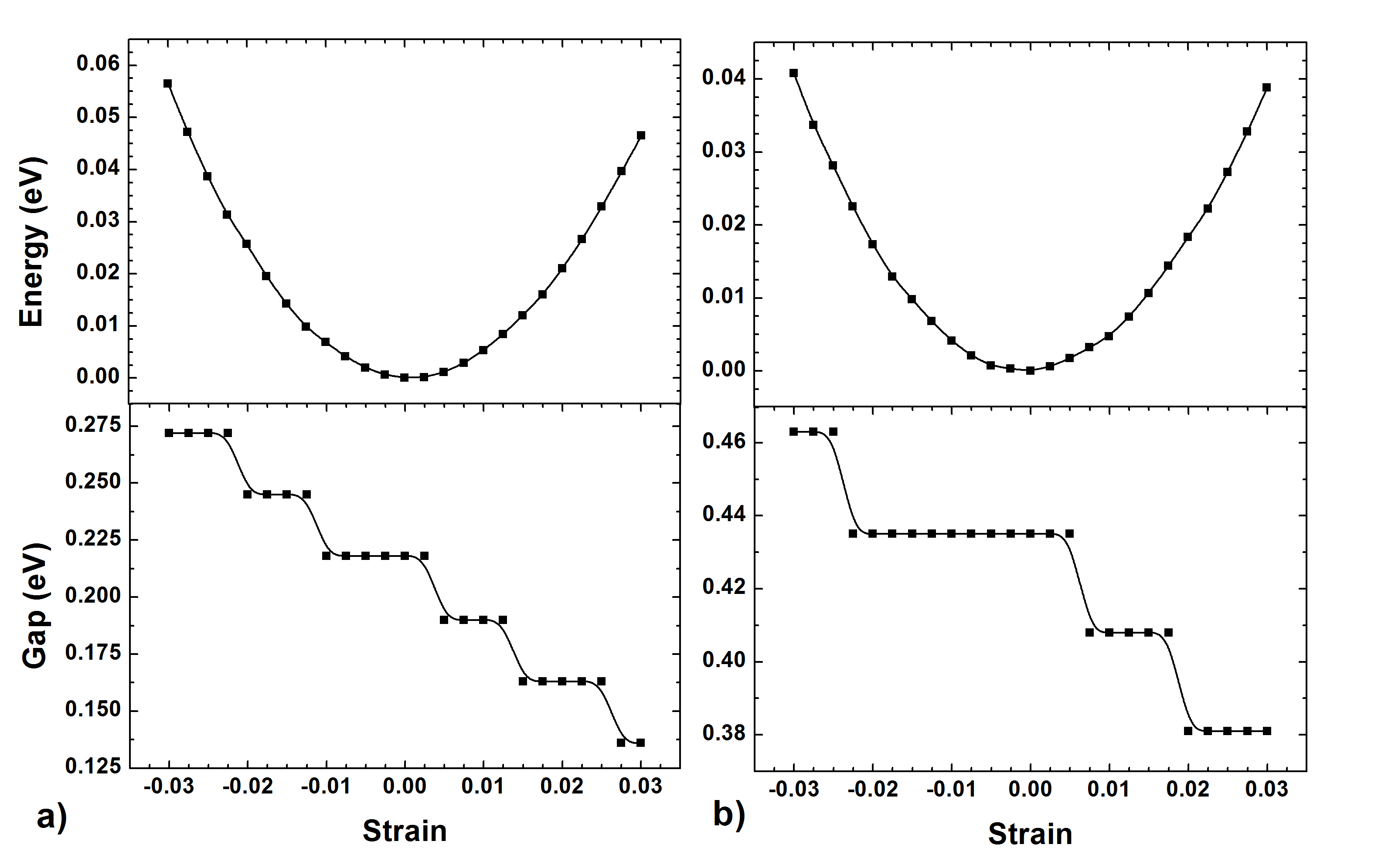}}
\caption{Total energy and bandgap value as a function of axial strain for (a) silicon and; (b) germanium nanotubes. The error in each gap value measurement is of 0.027 eV (approximately the thermal energy at room temperature), resulting in the shown plateaus.}
\label{ExStrain}
\end{figure*}

In order to investigate the mechanical properties of these tubes we studied their response to both tensile and compressive strains along their main axis direction. The energy versus strain curves are presented in Figure ~\ref{ExStrain}. We restrict ourselves to small strain values in order to maintain the system in the elastic region. In this regime, from the curves in Figure ~\ref{ExStrain} we can estimate the Young's modulus values for both structures by using

\begin{equation}
\label{YoungModulus}
Y=\frac{1}{V_{0}}\frac{\partial^2 U}{\partial \varepsilon ^2} ,
\end{equation}
where Y is the Young's modulus, $V_{0}$ is the equilibrium volume, U is the total strain energy and $\varepsilon$ is the strain. Using this equation we obtain values of $0.25TPa$ and $0.15TPa$ for the silicon and germanium tubes respectively. These values are quite high and comparable to the values found in strong metal alloys, although not as high as that of graphene, indicating these novel structures present promising mechanical properties. We have also calculated the Poisson's ratios along the directions shown in Figure ~\ref{figure1}. This ratio is defined by:

\begin{equation}
\label{PoissonRatio}
\nu_{i} =-\frac{\partial \varepsilon_{i}}{\partial \varepsilon_{a}} ,
\end{equation}
where $\nu_{i}$ is the Poisson ratio along the \emph{i} direction, $\varepsilon_{i}$ is the strain along the \emph{i} direction and $\varepsilon_{a}$ is the strain along the axial direction (assuming axial strain is applied). We obtained values of $\nu_{y}$ equal to 0.12 and 0.07 and $\nu_{z}$ equal to 0.12 and 0.06 for the silicon and germanium structures, respectively.

The electronic and mechanical properties are summarized in Table ~\ref{table1} along with the formation energy, which is defined as the energy per atom necessary for assembling the structure from isolate atoms. The formation energy values shown in Table \ref{table1} indicate that the tubular structures are more stable than their corresponding nanoribbons, which is in good accordance with the fact that silicon and germanium nanostructures favor $sp^3$ hybridizations over $sp^2$ ones. The nanoribbon structures had their geometry optmized following the exact same procedure applied to the tubular structures and presented structural buckling \cite{cahangirov2009two}. In order to have a conclusive proof of the stability of the tubular structures, the eigenvalues of the Hessian matrix were calculated. Considering that no negative values were found, we can affirm that they represent a local minimum \cite{banerjee1985search}, thus characterizing stable geometries \cite{baker1986algorithm}.

Concomitantly with the total energy analysis as a function of strain, we have also analysed the gap variation. The results are presented in Figure ~\ref{ExStrain}. The plateaus in the depicted curve are a consequence of the precision of our bandgap calculations of $0.001Ha$ (which is approximately $0.027eV$). As we can see from this figure, there are significant changes on the gap values for both structures as axial strain is applied. Both tubes undergo significant gap opening under compressive strain (negative values of horizontal axis) and gap closing under tensile strain (positive values of horizontal axis). Large gap changes of almost $50\%$ occur under strain rates as small as $3\%$. These changes at the gap values suggest possible applications of these novel structures as strain sensors and/or other applications where easy tuning of the gap is required.

\section{Summary and Conclusions}
\label{sec5}

In summary, we report the theoretical discovery of new silicon and germanium nanotubes which do not have carbon analogues due to the manifestation of the pseudo Jahn-Teller effect. These new structures do not present any helicity despite being neither armchair or zigzag. The tubes exhibit remarkable thermal and electronic properties. Silicon tubes are stable at temperatures as high as 500 K, while germanium ones can reach temperatures as high as 1000 K. Young's modulus values are considerably high at $0.25TPa$ and $0.15TPa$ and they have small indirect band gap values of $0.22eV$ and $0.44eV$, respectively. These band gap values can be significantly altered, for both structures, by compressive and tensile axial strains. Gap variations almost as large as $50\%$ for strain rates as small as $3\%$ were observed. Such properties could be extremely useful in the design of sensors and other technological devices. We believe these results can motivate new studies of silicon and germanium nanostructures, leading to searches beyond the limited spectrum of existing carbon nanostrucutres and, possibly, to the discovery of new unique and exciting materials.

\begin{section}{ACKNOWLEDGMENTS}
The authors thank Dr.  Felix  Hanke  and  Prof.  J.  A.  Brum  for helpful
discussions. We would especially like to thank  Prof. D. M. Ugarte for his
assistance with the TEM image calculations and very helpful discussions.
Work supported in part the the Brazilian  Agencies  CAPES,  CNPq  and
FAPESP.  The  authors acknowledge  the  Center  for  Computational
Engineering and  Sciences  at  Unicamp  for financial  support  through
the FAPESP/CEPID Grant \#2013/08293-7. RP acknowledges financial support of 
Fapesp Grant \#2011/17253-3. 

\end{section}

\end{document}